\documentclass[aps,prl,preprint,groupedaddress]{revtex4}
\usepackage{graphicx} \begin{document}

\title{Generalization of the Hartree-Fock-Bogoliubov theory: one and
two quasiparticle excitations }

\author{Paolo Tommasini, E. J. V. de Passos, M.O.C. Pires, A. F. R. de
Toledo Piza}

\affiliation{ Instituto de F\'{\i}sica, Universidade de S\~ao Paulo,
CP 66318\\ CEP 05389-970, S\~ao Paulo, \ SP, \ Brazil}

\date{\today} 

\begin{abstract}
 We present a generalization of the Hartree-Fock Bogoliubov (HFB)
theory in which the coupling between one and two quasi-particles
is taken into account.This is done by writing the excitation operators
as linear combinations of one and two HFB quasi-particles. The excitation
energies and the quasi-particle amplitudes are  given by  generalized 
Bogoliubov equations. The excitation spectrum has two branches. The
 first one is a discrete branch which is gapless and has a phonon
character at large wave-length and, contrarily to HFB, is always
stable. This branch is  detached from a continuum branch whose
threshold, at fixed total momentum, coincides with the two
quasi-particle threshold of the HFB theory. The gap between these 
two branches at P=0 is equal to two times the HFB gap, which then provides for
the  relevant energy scale. We also give  numerical results 
for a specific case.                         
\end{abstract}
 

\maketitle

\section{Introduction}

The experimental realization of Bose-Einstein condensation in trapped
neutral bosonic atoms has opened the opportunity for a comparison
 of microscopic theories of dilute systems with  experimental data \cite{1}.
The standard approach is to solve the Gross-Pitaevskii (GP) equation for the condensate wave function 
and the linear Bogoliubov-de Gennes  
(BdG) equations for the collective excitations \cite{2}. For $T=0$ this 
theory has
been successfully compared with the existing data \cite{3}-\cite{5}. \\
Physically, the GP + BdG theory is a free quasi-particles theory.
However it is not the best of such theory from a
variational point of view, being superseded by the Hartree-Fock- 
Bogoliubov (HFB) theory \cite{6}, where the non-linear
BdG equations are solved self-consistently and the GP and BdG
equations are coupled. The two approaches agree when we neglect the
depletion of the condensate in the self-consistent theory. 
This would suggest that  HFB  is the proper theory
to use when the fluctuations become important, for example, 
in trapped gases through the
mechanism of Feshbach resonance \cite{7} or at finite
temperature. It is however 
 well known that the HFB theory has problems when applied to
homogeneous systems \cite{6}. Indeed, the excitation spectrum has a gap which
violates the Hugenholtz-Pines theorem, which states that the excitation
spectrum should be gapless \cite{8}. The HFB theory is also in 
contradiction with some thermodynamic data in $^{4}$He 
(for example, the specific heat),
which requires that, in the large wave-length limit, the excitation
spectrum should have a phonon behavior \cite{9}.\\

In this paper we present a generalization of the HFB theory leading to
an excitation spectrum which, in particular, eliminates the gap
problem. The key ingredient for this is the inclusion of two
quasi-particle components together with their coupling to one
quasi-particle components for the description of the excitation
spectrum. Excitation energies and the structure of the corresponding 
modes are given by generalized Bogoliubov equations. The excitation spectrum
includes a discrete stable phonon-like excitation branch. In addition
to this discrete branch there is a continuum branch whose threshold
is located at twice the HFB quasi-particle energy. This branch has,
therefore, a gap at total momentum ${\bf P}=0$ which is twice the HFB
gap. Therefore, this gap sets an important energy scale for the excitation
spectrum. \\

The use of this mechanism to solve the HFB gap problem has in fact
being pioneered more than for decades ago by Takano \cite{10}. A more
recent work by Hutchinson at al \cite{11} deals with the coupling
between one and two quasi-particle components in a perturbative way,
while the work of Kerman and Tommasini \cite{12} deals with the same problem on
the basis of the Gaussian  functional approximation to a field
theoretical variational procedure. \\

In this paper we use standard equations of motion techniques
\cite{14,15}, employed, for example, in reference \cite{13} to study
chiral symmetry restoration in the linear $\sigma$ model
non-pertubatively. This allows for a clear identification of the
dynamical role played by various parts of the many-body hamiltonian
when expanded in terms of quasi-particles. \\

Our paper is organized as follows: In section 2 we briefly discuss the
basic properties of the HFB theory. In section 3 we derive the
generalized Bogoliubov equations for the excitation energies and
quasi-particle amplitudes. Our derivation allows for a clear
identification of parts of the hamiltonian responsible for the
coupling between on and two quasi-particle components of the
excitation operator. In section 4, we show that there exist 
a Goldstone mode at momentum equal to zero. Our proof of its
existence is very simple and clearly related  to the violation of
number conservation. The results of a numerical
application  of the theory are discussed in section
5. Specifically we examine the properties of the excitation spectrum,
its stability, and change in physical content as a function of the
total momentum ${\bf P}$. We also make a comparison with the HFB and
Bogoliubov approximations. Our conclusions are presented in section 6. 
All the expressions needed for  numerical applications are given in 
appendices $A$ and $B$.

\section{The HFB theory}
The starting point is the Grand-Hamiltonian written in second quantization as 

\begin{equation}
\hat{h} = \hat{H} - \mu \hat{N} = \sum_{\bf k} (e_{k} - \mu) a_{\bf k}^{\dagger} 
a_{\bf k} + \frac{1}{2} \sum_{\bf k_{1}, k_{2},{\bf q}}  V(q) a_{{\bf k_{1}} + 
{\bf q}}^{\dagger} a_{{\bf k_{2}}-{\bf q}}^{\dagger} a_{\bf k_{1}} a_{\bf k_{2}}
\label{hamiltonian}
\end{equation}

\noindent where $e_{k}$ is the free particle kinetic energy,
\begin{equation}
e_{k} = \frac{\hbar^{2} k^{2}}{2m}, 
\end{equation}

\noindent  $V(q)$ is the Fourier transform per unit volume  of the
atom-atom interaction potential

\begin{equation}
V(q) = \frac{1}{\Gamma} \int V(r) e^{i {\bf q}.{\bf r}} d^{3}r
=\frac{\tilde{V}(q)}{\Gamma},
\label{fourier}
\end{equation}

\noindent and the operators $a_{\bf k}^{\dagger}$ e $a_{\bf k}$, respectively, 
create and annihilate atoms in a state with momentum $\hbar {\bf k}$, 
the corresponding  wave-function $\exp(i {\bf k}.{\bf
r})/\sqrt{\Gamma}$  
satisfying periodic boundary conditions in a cubic box of volume $\Gamma$. \\

In a first step we perform a canonical transformation to the quasi-particles by 
introducing a new set of creation and annihilation operators through the 
 Bogoliubov rotation \cite{16}

\begin{equation}
a_{\bf k} =c_{\bf k} + z_{0} \delta_{{\bf k},0}= u_{\bf k} \eta_{\bf k} 
- v_{\bf k} \eta_{-\bf k}^{\dagger} + 
z_{0} \delta_{{\bf k},0}, 
\label{bogoliubov}
\end{equation}

\noindent where $u_{\bf k}$ and $v_{\bf k}$ are even functions of ${\bf k}$, 
$u_{\bf k} = u_{-{\bf k}}$, $v_{\bf k} = v_{-{\bf k}}$, and  $z_{0}$ is a c-number. 
The constant $z_{0}$  appears as a shift in the equation for ${\bf k}=0$ to account 
for the macroscopic condensate in the zero momentum state. In order to render the 
transformation canonical, the Bogoliubov factors have to obey 
the constraint

\begin{equation}
u_{\bf k}^{2} - v_{\bf k}^{2} = 1. 
\label{canonical}
\end{equation}

\noindent It is straightforward to write 
the Grand-Hamiltonian $\hat{h}$ in the 
quasi-particle basis. After normal ordering one obtains,
\begin{equation}
\hat{h} = \hat{H} - \mu \hat{N} = h_{0}+ \hat{h}_{1} +\hat{h}_{2}+
 \hat{h}_{3} + \hat{h}_{4}
\label{normalhamiltonian}
\end{equation}

\noindent where the normal ordered operators $\hat{h_{i}}$ contain
 $i$ quasi-particles. They are written explicitly in  Appendix A. \\

 The amplitudes $u_{\bf k}$, $v_{\bf k}$ and the
shift $z_{0}$ are determined in the HFB theory by 
minimizing the expectation value of 
$\hat{h}$, $\langle \Phi | \hat{h} | \Phi \rangle$ 
in the quasi-particle vacuum, that is, $\eta_{\bf k} |\Phi \rangle
=0$. Since 
the only term which contributes to the expectation value is the 
term $h_{0}$, the minimization is 
equivalent to the equations

\begin{eqnarray}
\frac{\partial h_{0}}{\partial z_{0}} &=& 0  \label{equilibriumz0}\\
\frac{\partial h_{0}}{\partial v_{\bf k}} + \frac{\partial h_{0}}{\partial u_{\bf k}} 
\frac{\partial u_{\bf k}}{\partial v_{\bf k}} &=& 0
\label{equilibriumuv}
\end{eqnarray}

\noindent with $h_{0}$ given by

\begin{eqnarray}
h_{0} &=& - z_{0}^{2} \mu + \frac{z_{0}^{4}}{2} V(0) + \sum_{\bf k} 
[e_{k} - \mu + ( V(0) + V(k)) z_{0}^{2}] v_{\bf k}^{2} \nonumber \\
&& - \sum_{\bf k} V(k) z_{0}^{2} u_{\bf k} v_{\bf k} + \frac{1}{2} 
\sum_{{\bf k_{1}},{\bf k_{2}}} (V(0) + V(|{\bf k_{1}} - {\bf k_{2}}|) 
v_{\bf k_{1}}^{2} v_{\bf k_{2}}^{2} \nonumber \\
&& + \frac{1}{2} \sum_{{\bf k_{1}},{\bf k_{2}}} V(|{\bf k_{1}}
 - {\bf k_{2}}|) u_{\bf k_{1}} v_{\bf k_{1}} u_{\bf k_{2}} v_{\bf k_{2}}
\label{H0}
\end{eqnarray} 

The two equations above can be written in a very compact way if we introduce the 
Hartree, exchange and pair potentials defined by the following
relations \cite{17}:

\begin{eqnarray}
{U}_{h} &=& \sum_{\bf k_{1}} V(0) 
\langle \Phi | a_{\bf k_{1}}^{\dagger} a_{\bf k_{1}} | \Phi \rangle \label{Potencials1} \\
 {U}_{ex}(\bf k) &=& \sum_{\bf k_{1}} V(|{\bf k} - {\bf k_{1}}|)
 \langle \Phi | a_{\bf k_{1}}^{\dagger} a_{\bf k_{1}} | \Phi \rangle \label{Potencials2} \\
{U}_{pair}(\bf k) &=& \sum_{\bf k_{1}} V(|{\bf k} - {\bf k_{1}}|)
 \langle \Phi | a_{\bf k_{1}} a_{-\bf k_{1}} | \Phi \rangle 
\label{Potencials3}
\end{eqnarray}

\noindent These potentials can be written as the sum of two terms which can be 
interpreted as related to the condensate and to the non-condensate, respectively

\begin{eqnarray}
{U}_{h} &=&   U_{h}^{c} +  U_{h}^{nc} = V(0) z_{0}^{2} + \sum_{\bf k_{1}} V(0) 
\langle \Phi | c_{\bf k_{1}}^{\dagger} c_{\bf k_{1}} | \Phi \rangle \\
{U}_{ex}(\bf k) &=& U_{ex}^{c} +  U_{ex}^{nc} = V(k) z_{0}^{2}
 + \sum_{\bf k_{1}} V(|{\bf k} - {\bf k_{1}}|)
 \langle \Phi | c_{\bf k_{1}}^{\dagger} c_{\bf k_{1}} | \Phi \rangle \\
{U}_{pair}(\bf k) &=& U_{pair}^{c} +  U_{pair}^{nc} = V(k) z_{0}^{2} 
+  \sum_{\bf k_{1}} V(|{\bf k} - {\bf k_{1}}|)
 \langle \Phi | c_{\bf k_{1}} c_{-\bf k_{1}} | \Phi \rangle 
\end{eqnarray}

\noindent In terms of these potentials the equilibrium equations (\ref{equilibriumz0}) 
and (\ref{equilibriumuv}) can be written as

\begin{eqnarray}
&& z_{0} [-\mu + {U}_{h} + {U}_{ex}^{nc}(0) + {U}_{pair}^{nc}(0)] = 0 \label{equilibrium1} \\
&& \tanh 2 \sigma_{k} = \frac{{U}_{pair}(\bf k)}{e_{k} + {U}_{h} + {U}_{ex}(\bf k) - \mu} = 
\frac{{U}_{pair}(\bf k)}{\tilde{e}(\bf k)} \label{equilibrium2},
\end{eqnarray}

\noindent with $u_{\bf k} = \cosh \sigma_{\bf k}$ and $v_{\bf k}= \sinh \sigma_{\bf k}$. \\

The quasi-particle vacuum $|\Phi \rangle$ does not have a definite number of 
particles. In order to control the number of particles we determine $\mu$ from
the condition that the mean value of the number of particles in the state 
$|\Phi \rangle$ is $N$, $N = \langle \Phi| \hat{N} | \Phi \rangle$ which gives  

\begin{equation}
N = z_{0}^{2} + \sum_{\bf k} v_{\bf k}^{2}
\label{constraint}
\end{equation}

\noindent Thus, the set of equations (\ref{equilibrium1}), (\ref{equilibrium2}) 
and (\ref{constraint}) determine $\mu$, $z_{0}$ and the Bogoliubov amplitudes 
$u_{\bf k}$ and $v_{\bf k}$. Equations ( \ref{equilibrium1}) and 
(\ref{equilibrium2}) 
can also be derived by demanding that $\hat{h}_{1}$ vanishes and that 
$\hat{h}_{2}$ is diagonal 
in the quasi-particle basis, $\hat{h}_{2} = \sum_{\bf k} \omega({\bf
k}) \eta_{\bf k}^{\dagger} \eta_{\bf k}$,
with $\omega({\bf k})$ the quasi-particle energies, 
$\omega({\bf k})^{2} = \tilde{e}({\bf k})^{2} -{U}_{pair}(\bf k)^{2}$. \\

One feature of the HFB theory is that the excitation energy have a
gap in the 
limit $k \rightarrow 0$ \cite{6},  

\begin{equation}
\omega(0)^{2} = - 4 \tilde{V}(0) n_{0} {U}_{pair}^{nc}(0)
\label{gap}
\end{equation}
\noindent where  $n_{0}= z_{0}^{2}/\Gamma$ is the condensate
density. The 
existence of an energy gap in the excitation spectrum does not agree with 
the phonon spectrum in superfluid systems and is also in contradiction
with 
the  Hugenholtz-Pines (HP) \cite{8} theorem, which sates that the 
energy of an excitation with wavenumber ${\bf k}$  of a 
many-body system should vanish when $k \rightarrow 0$. \\ 

An approximate way to satisfy the HP theorem is to neglect the so-called anomalous 
density $m_{\bf k} = \langle \Phi | c_{\bf k} c_{-\bf k} | \Phi \rangle$ in the 
HFB theory \cite{6}. In this approximation $U_{pair}^{nc}(\bf k)$
vanishes and the gap disappears. This approximation is known as the 
Popov approximation. In the next section we are going to present a
theory that has a gapless dispersion 
relation while taking $m_{\bf k}$ fully into account. As it turns out,
this theory also gives a physical meaning to (\ref{gap}).

\section{The quasi-particle RPA} 

As is well known in many-body physics the Random Phase Approximation
(RPA) singles out the Goldstone mode due to a symmetry breaking at the
mean field level \cite{14,15}.

One among the many ways of  deriving  the RPA equations, is the
linearization of the equations of motion \cite{21}. In principle, if we
find operators satisfying the equations\\
\begin{eqnarray} 
[H,Q_{\alpha}^{\dagger}] &=& \Omega_{\alpha} Q_{\alpha}^{\dagger}
\label{RPA1}
\\
Q_{\alpha} | \Psi_{0} \rangle &=& 0 
\label{RPA2}
\end{eqnarray}
with the normalization condition \\
\begin{equation} 
\langle \Psi_{0} | [ Q_{\beta},Q_{\alpha}^{\dagger}]
|\Psi_{0}\rangle = \delta_{\beta , \alpha} 
\end{equation}
where $ |\Psi_{0} \rangle $ is the exact ground state, we have found
 an exact excited state of  the
many-body system since, from the above equations, it follows that
$|\Psi_{\alpha} \rangle =Q_{\alpha}^{\dagger} | \Psi_{0} \rangle $
is an eigenstate of H with excitation energy equal to $
\Omega_{\alpha}$. However this cannot be carried out in general, and
 we are bound to use approximate methods regarding both $Q_{\alpha}$
 and $| \Psi_{0} \rangle$  in the solution of 
equations (\ref{RPA1}) and
(\ref{RPA2}). In the method of the linearization of the equations 
of motion we make an ansatz about the form of the excitation
operators, writing it as a 
linear combination of basic excitations and we linearize the left
-hand side of equation (\ref{RPA1}) with respect to these operators.

In this paper we look for excitation operators which are a combination of 
one and two HFB quasi-particles,
\begin{equation}
Q_{\bf P}^{\dagger} = x_{\bf P} \eta_{\bf P}^{\dagger} + 
y_{\bf P} \eta_{-\bf P} + \sum_{{\bf q} \ge 0} X_{{\bf q},{\bf P}} 
\frac{\eta_{{\bf q} + {\bf P}/2}^{\dagger} \eta_{-{\bf q} + 
{\bf P}/2}^{\dagger}}{\sqrt{1 + \delta_{{\bf q},0}}} +   
Y_{{\bf q},{\bf P}} 
\frac{\eta_{-{\bf q} - {\bf P}/2} \eta_{{\bf q} - 
{\bf P}/2}}{\sqrt{1 + \delta_{{\bf q},0}}}
\label{newquasiparticles}
\end{equation}  

In the equation (\ref{newquasiparticles}) $Q_{\bf P}^{\dagger}$ creates an
excitation with momentum ${\bf P}$ and it is a linear combination of
one, $\eta_{\bf P}^{\dagger}$,
$ \eta_{-\bf P}$ , and two, $\eta_{{\bf q} + {\bf P}/2}^{\dagger} 
\eta_{-{\bf q} + 
{\bf P}/2}^{\dagger} $, $\eta_{-{\bf q} - {\bf P}/2} \eta_{{\bf q} - 
{\bf P}/2}$ HFB quasi- particles. This last term creates (annihilates)
a pair with 
total momentum ${\bf P}(-{\bf P})$ and relative momentum $\bf q$.
The coefficients 
$x_{\bf P}$, $y_{\bf P}$, $X_{{\bf q},{\bf P}}$ and $Y_{{\bf q},{\bf P}}$ 
are even functions of $\bf P$. As the pair creation and annihilation
are invariant by
the replacement ${\bf q} \rightarrow -{\bf q}$, $X_{{\bf q},{\bf P}}$
and $Y_{{\bf q},{\bf P}}$
are even functions of $\bf q$ and we restrict the sum in order that
the pairs appear only once. \\

The coefficients in (\ref{newquasiparticles}) are determined  by the method of 
 the equations of motion  in the version of  references \cite{14}and
\cite{15}, which is a systematic way of achieving the linearization
referred above. Following these references, notice that from equations
 (\ref{RPA1}) and (\ref{RPA2}) one has

\begin{equation}
\langle\Psi_{0} |[  Q_{\bf P}, H , Q_{\bf P}^{\dagger}]| \Psi_{0} \rangle = 
\Omega_{P} \langle \Psi_{0} |[ Q_{\bf p}, Q_{\bf P}^{\dagger}]|\Psi \rangle
\label{RPAl}
\end{equation}
\noindent where $[A,B,C]$ is the  symmetrized double 
commutator $\frac{1}{2} ([A,[B,C]] + [[A,B],C])$.

Requiring that $\Omega_{\bf P}$ be stationary in a variation of the
excitation operators one has: 
\begin{equation}
\langle\Phi |[ \delta Q_{\bf P}, H , Q_{\bf P}^{\dagger}]| \Phi \rangle = 
\Omega_{P} \langle \Phi |[\delta Q_{\bf p}, Q_{\bf P}^{\dagger}]|\Phi \rangle
\label{RPA}
\end{equation}
where $\delta Q_{\bf P} $ is the hermitian conjugate of the operator given
by the variation  of the coefficients in (\ref{newquasiparticles}) and,
as usual, we replaced the ground state $|\Psi_{0} \rangle $ by the HFB
vacuum , $|\Phi \rangle $. 

Performing the variation indicated in equation (\ref{RPA}), 
 we get for the excitation energies 
$\Omega_{P}$ and the coefficients
\begin{equation}
{\cal X} = \left( \begin{array}{c}
              x_{\bf P}  \\
              X_{\bf P}        
                    \end{array} \right) \quad \mbox{and} \quad
{\cal Y} = \left( \begin{array}{c}
              y_{\bf P}  \\
              Y_{\bf P}        
                    \end{array} \right),  \label{nono}
\end{equation}
\noindent the equation

\begin{equation} 
\left(  \begin{array}{cc}
       {\cal A} &   {\cal B} \\      
  {{\cal B}^{\ast}}  &   {\cal A}^{\ast}
       \end{array}
\right) 
\left(  \begin{array}{c}        
       { \cal X } \\
         { \cal Y} 
       \end{array}
\right) =
\Omega_{P}
\left(
\begin{array}{cc}
       1  &  0 \\
       0  & -1
\end{array}    
\right)
\left(\begin{array}{c}        
      { \cal X} \\
       {\cal Y} 
       
       \end{array}  
\right ),
\label{matrix}
\end{equation}
 with the coefficients obeying the  normalization condition
\begin{equation}
\langle \Phi | [ Q_{\bf P}^{\lambda},Q_{\bf P}^{\dagger^{\tau}}] 
\Phi \rangle =  x_{\bf P}^{\lambda^{\ast}} x_{\bf P}^{\tau} -
y_{\bf P}^{\lambda^{\ast}} y_{\bf P}^{\tau} + \sum_{{\bf q} \ge 0} 
\left( X_{{\bf q},{\bf P}}^{\lambda^{\ast}}  
X_{{\bf q},{\bf P}}^{\tau}- Y_{{\bf q},{\bf P}}^{\lambda^{\ast}} 
Y_{{\bf q},{\bf P}}^{\tau} \right)=\delta_{\lambda,\tau}. 
\label{normalization}
\end{equation}
\noindent For each {\bf P} we have a number of modes 
equal to the number of operator pairs plus one, 
$n_{\mbox pairs} +1$, actually this number is  
denumerable infinite,  and this is denoted by the quantum
numbers $\lambda$, $\tau$ in equation (\ref{normalization}). \\

In expression (\ref{nono}), $X_{\bf P} (Y_{\bf P})$ stands for the 
set of $n_{\mbox pairs}$ coefficients $X_{{\bf q},{\bf P}}(Y_{{\bf
q},{\bf P}})$ and ${\cal A}$ and ${\cal B}$ 
are, respectively, hermitian and symmetric 
matrices of dimension $n_{\mbox pairs} + 1$. \\

The hermitian matrix ${\cal A}$ is given by
\begin{equation}
{\cal A} = \left(\begin{array}{cc}
           {\cal A}^{11} & {\cal A}^{12} \\
           {\cal A}^{21} &  {\cal A}^{22}
           \end{array}       
\right).
\end{equation}
          
\noindent The diagonal blocks  ${\cal A}^{11}$ and  ${\cal A}^{22}$
are hermitian matrices with dimensionality 1 and $n_{\mbox pairs}$ 
respectively. They are given by  

\begin{equation}
{\cal A}^{11}({\bf P}) = \langle \Phi | [ \eta_{\bf P}, H, \eta_{\bf P}^{\dagger}] | \Phi \rangle,
\end{equation}
\noindent and 

\begin{equation}
{\cal A}^{22}({\bf q'},{\bf q};{\bf P}) = \langle \Phi |[ \frac{\eta_{{\bf q'} + 
{\bf P}/2} \eta_{-{\bf q'} + {\bf P}/2}}{\sqrt{1+ \delta_{{\bf q'},0}}}, H , \frac{\eta_{{\bf q} + 
{\bf P}/2}^{\dagger} \eta_{-{\bf q} + {\bf P}/2}^{\dagger}}{\sqrt{1+ \delta_{{\bf q},0}}} | \Phi \rangle.
\end{equation}

\noindent The coupling matrices  ${\cal A}^{12}$ and  ${\cal A}^{21}$
are hermitian conjugates with dimensions  $1 \times n_{\mbox pairs}$
and $ n_{\mbox pairs} \times 1$ respectively and are given by

\begin{equation}
{\cal A}^{12}({\bf q}; {\bf P}) = \langle \Phi | [ \eta_{\bf P}, H,  \frac{\eta_{{\bf q} + 
{\bf P}/2}^{\dagger} \eta_{-{\bf q} + {\bf P}/2}^{\dagger}}{\sqrt{1+ \delta_{{\bf q},0}}}  ] | \Phi \rangle,
\end{equation}

The symmetric matrix ${\cal B}$ can be split in a similar way

\begin{equation}
{\cal B} = \left(\begin{array}{cc}
           {\cal B}^{11} & {\cal B}^{12} \\
           {\cal B}^{21} &  {\cal B}^{22}
           \end{array}       
\right).
\end{equation}

\noindent The diagonal blocks  ${\cal B}^{11}$ and ${\cal B}^{22}$ are
symmetric matrices whose elements are given by

\begin{equation}
{\cal B}^{11}({\bf P}) = \langle \Phi | [ \eta_{\bf P}, H, \eta_{\bf -P}] | \Phi \rangle,
\end{equation}

 \noindent and

\begin{equation}
{\cal B}^{22}({\bf q'},{\bf q};{\bf P}) = \langle \Phi |[ \frac{\eta_{{\bf q'} + 
{\bf P}/2} \eta_{-{\bf q'} + {\bf P}/2}}{\sqrt{1+ \delta_{{\bf
q'},0}}}, H , \frac{\eta_{-{\bf q} - 
{\bf P}/2} \eta_{{\bf q} - {\bf P}/2}}{\sqrt{1+ \delta_{{\bf q},0}}} | \Phi \rangle.
\end{equation}

\noindent The coupling matrices ${\cal B}^{12}$ and  ${\cal B}^{21}$
are transpose of each other and given by 

\begin{equation}
{\cal B}^{12}({\bf q}; {\bf P}) = \langle \Phi | [ \eta_{\bf P}, H,  \frac{\eta_{-{\bf q} - 
{\bf P}/2} \eta_{{\bf q} - {\bf P}/2}}{\sqrt{1+ \delta_{{\bf q},0}}}  ] | \Phi \rangle,
\end{equation}

Note that the matrices 12 and 21 couple one and two quasi-particle excitations 
whereas the matrices 11 and 22 act only inside the one and two
quasi-particle basis, respectively. As shown in appendix B, the
matrices 12 and 21 depend only on $\hat{h}_{3}$, (see
Eq. 6). Therefore 
this term is responsible for the coupling between the one and two 
quasi-particle components. On the other hand the matrices 11 and 22,
that do not mix these two components, depend only on $\hat{h}_{2}$ and
$\hat{h}_{4}$. All 
the matrix elements are computed in detail in  appendix B.\\

If the coupling terms $\hat{h}_{3}$ and $\hat{h}_{4}$ are
set to zero we have that ${\cal B}=0$ and ${\cal A} $ is diagonal
with eigenvalues  $\omega({\bf P})$ and $\omega_{2}({\bf q},{\bf P}) =
\omega({\bf q}+{\bf P}/2)+ \omega({-\bf q}+{\bf P}/2)$
which correspond to one and two free quasi-particle energies.\\

As the static quantities are real, the elements of the matrices ${\cal
A}$ and ${\cal B} $ are real and the  RPA equations can be written 
in a more compact and symmetrical form if we introduce the following 
new variables:

\begin{eqnarray}
\phi_{1}({\bf P}) &=& x_{\bf P} + y_{\bf P} \nonumber \\
\phi_{2}({\bf q};{\bf P}) &=& X_{{\bf q},{\bf P}} + Y_{{\bf q},{\bf P}} \nonumber \\
\pi_{1}({\bf P}) &=& ({x_{\bf P} - y_{\bf P}}) \nonumber \\
\pi_{2}({\bf q};{\bf P}) &=& ({X_{{\bf q},{\bf P}} - Y_{{\bf q},{\bf P}}})  
\end{eqnarray}

\noindent Grouping $\phi_{1}$,$\phi_{2}$ and $\pi_{1}$, $\pi_{2}$ in order 
to have the following  $n_{\mbox{pairs}}+1 $ column matrices

\begin{equation}
{\cal Q} = \left(\begin{array}{c}
           \phi_{1}({\bf P}) \\
           \phi_{2}({\bf q};{\bf P})
           \end{array}       
\right)
\;\;\;\;\;\
{\cal P} = \left(\begin{array}{c}
           \pi_{1}({\bf P}) \\
           \pi_{2}({\bf q};{\bf P})
           \end{array}       
\right)
\label{QPRPA}
\end{equation}

\noindent we can rewrite the equations (\ref{matrix}) in a very
compact form \cite{11,12}

\begin{eqnarray}
 \Omega {\cal Q}  &=& {\cal A}_{-}{ \cal P}  \label{OmegaQ}\\
 \Omega {\cal P} &=& {\cal A}_{+} {\cal Q}   \label{OmegaP}
\end{eqnarray}

\noindent where now we work with two symmetric matrices ${\cal A}_{+}$
and ${\cal A}_{-}$  which are given in terms of the matrices ${\cal
A}$ and ${\cal B}$ as 

\begin{eqnarray}
{\cal A}_{+} &=& {\cal A} + {\cal B} \label{A+} \\
{\cal A}_{-} &=& {\cal A}-{\cal B} \label{A-}
\end{eqnarray}

\noindent The elements of these matrices can be written as:

 \begin{eqnarray}
&& {\cal A}_{-}^{11} = {\cal A}_{+}^{11}= \omega({\bf P}) \\
&& {\cal A}_{-}^{12} ({\bf q},{\bf P}) =  \frac{z_{0}}{\sqrt{1 + \delta_{{\bf q},0}}} C_{0}({\bf P}) \{ C_{0}(-) C_{0}(+) 
 [ V(+)+ V(-)] + [C_{0}(+) C_{0}(-) \nonumber \\
&& - C_{0}^{-1}(+) C_{0}^{-1}(-)] V({\bf P}) \}  \\ 
&& {\cal A}_{+}^{12}({\bf q},{\bf P})  = \frac{z_{0}}{2\sqrt{1 +
\delta_{{\bf q},0}}} C_{0}^{-1}({\bf P})
 \{ [C_{0}^{-1}(+) C_{0}(-) + C_{0}(+) C_{0}^{-1}(-)][V(+) +  \nonumber \\ 
&&  V(-)]  \}  \\
&& {\cal A}_{-}^{22}({\bf q},{\bf q}',{\bf P}) =  
[\omega(+)+\omega(-)]\delta_{{\bf q},{\bf q}'} + \nonumber \\
&&\frac{1}{2 \sqrt{(1+ \delta_{{\bf q},0})(1+ \delta_{{\bf q}',0})}}
 \{ [C_{0}^{-1}(+') C_{0}^{-1}(-') C_{0}^{-1}(+)C_{0}^{-1}(-) \nonumber \\
&& + C_{0}(+') C_{0}(-') C_{0}(+)C_{0}(-)][V(|{\bf q}-{\bf q}'|)   
 + V(|{\bf q}+{\bf q}')] + [C_{0}^{-1}(+') C_{0}^{-1}(-')- \nonumber \\ 
&& C_{0}(+') C_{0}(-')][C_{0}^{-1}(+) C_{0}^{-1}(-)- C_{0}(+)
C_{0}(-)] V({\bf P}) \}  \\ 
&& {\cal A}_{+}^{22}({\bf q},{\bf q}',{\bf P}) =
[\omega(+)+\omega(-)] \delta_{{\bf q},{\bf q}'} + \nonumber \\
&&\frac{1}{2 \sqrt{(1+ \delta_{{\bf q},0})(1+ \delta_{{\bf q}',0})}} 
[C_{0}^{-1}(+') C_{0}(-') C_{0}^{-1}(+) C_{0}(-)+ \nonumber \\ 
&& C_{0}(+') C_{0}^{-1}(-') C_{0}(+) C_{0}^{-1}(-)]V(|{\bf q}-{\bf q}'|)+[ C_{0}^{-1}(+') C_{0}(-') C_{0}^{-1}(-) C_{0}(+)+ \nonumber \\ 
&& C_{0}(+') C_{0}^{-1}(-') C_{0}(-) C_{0}^{-1}(+)] V(|{\bf q}+{\bf q}'|). 
\end{eqnarray}

\noindent  These expressions show the remarkable
result that all the matrix elements depend on the static factors
through only one function i.e. 

\begin{equation}
C_{0}({\bf q}) = u_{\bf q} - v_{\bf q}
\end{equation}

\noindent where we used the notation $\pm= \pm {\bf q} +{\bf P}/2$ and
$\pm'=\pm {\bf q}' + {\bf P}/2$. By solving the coupled equations (\ref{OmegaQ}) and  (\ref{OmegaP}) we will find the new 
excitations which  will now have one and two quasi-particle contributions. The
results for the excitation energies will lead to a  discrete
branch and a continuum whose threshold coincides with the two 
quasi-particle threshold of the  HFB theory.  The discrete branch 
is detached from the continuum and due to the coupling between the one
and two quasi-particles will be pushed down and become gapless. This 
fact can be proved for any   pseudo-potential as will be shown in 
the next section

\section{The Goldstone Mode}

Equations (\ref{matrix}) have a zero energy solution when ${\bf P}=0$, 
$\Omega_{0}=0$, with zero-norm, the Goldstone mode \cite{18}. To identify this 
solution one has to consider the generator of the symmetry violated by the theory 
which, in our case, is the $U(1)$ symmetry whose generator is the 
number operator. This operator when written in the HFB basis possesses
components that are present in the general "anzatz" for the excitation
operators, eq.(\ref{newquasiparticles}). These components will be identified 
with the excitation operator of the Goldstone mode $\hat{Q}_{G}$. \\

Thus to find  $\hat{Q}_{G}$ we start writing  the number operator

\begin{equation}
\hat{N} = \sum_{\bf k} a_{\bf k}^{\dagger} a_{\bf k}
\label{N}
\end{equation}

\noindent in terms of the HFB quasi-particles $\eta_{\bf k}^{\dagger}$, 
$\eta_{\bf k}$, giving

\begin{equation}
\hat{N} = N + z_{0}(u_{0}-v_{0})(\eta_{0} + \eta_{0}^{\dagger})
 - \sum_{{\bf k}\ge 0} \frac{2 u_{\bf k} v_{\bf k}}{1 + \delta_{{\bf k},0}} 
( \eta_{\bf k}^{\dagger} \eta_{-{\bf k}}^{\dagger} +  
\eta_{\bf k} \eta_{-{\bf k}}) +\sum_{\bf k} (u_{\bf k}^{2} + v_{\bf k}^{2}) \eta_{\bf k}^{\dagger} \eta_{\bf k}. 
\label{Nhat}
\end{equation}

\noindent Comparing with the general "anzatz" eq.(\ref{newquasiparticles}) 
we identify the excitation operator of the Goldstone mode with

\begin{equation}
\hat{Q}_{G} = z_{0}(u_{0}-v_{0})(\eta_{0} + \eta_{0}^{\dagger}) - 
\sum_{{\bf k}\ge 0} \frac{2 u_{\bf k} v_{\bf k}}{1 + \delta_{{\bf k},0}} 
( \eta_{\bf k}^{\dagger} \eta_{-{\bf k}}^{\dagger} +  
\eta_{\bf k} \eta_{-{\bf k}}).
\end{equation}

\noindent Given this form,  our next task is to prove that, when ${\bf P}=0$, 
there is a zero-energy solution of (\ref{matrix}) of zero norm with,

\begin{eqnarray}
x_{0} &=& y_{0} = z_{0}(u_{0}-v_{0}) \\
X_{{\bf q},0} &=& Y_{{\bf q},0} = - \frac{2 u_{\bf q} v_{\bf q}}
{\sqrt{1 + \delta_{{\bf q},0}}}.
\end{eqnarray}

 Equation (\ref{matrix}) is equivalent to the coupled 
equations (\ref{OmegaQ}) and (\ref{OmegaP}). Since the Goldstone mode
has zero-energy and zero-norm one has $\Omega_{0}=0$ and ${\cal P}=0$ 
and the coupled equations reduce to

\begin{equation}
{\cal A}_{+} {\cal Q}_{G}=0 
\end{equation}

\noindent with

\begin{equation}
{\cal Q}_{G} = \left(\begin{array}{c}
           \phi_{1}(0) \\
           \phi_{2}({\bf q};0)
           \end{array}       
\right) = \left(\begin{array}{c}
           2 x_{0} \\
           2 X_{{\bf q},0}
           \end{array}       
\right)
\end{equation}

\noindent From the expression of the matrix elements of ${\cal A}_{+}$
, Eqs. (44) and (46) (at ${\bf P}=0$), ${\cal Q}_{G}$ given above is a
 zero-energy solution of 
the eqs. (\ref{OmegaQ}) and (\ref{OmegaP}) provided the
following identities are satisfied

\begin{eqnarray}
&& \frac{1}{2} \omega_{0} ( u_{0}-v_{0})^{2} + U_{pair}^{nc}(0) = 0  \\
&& U_{pair}({\bf q}) - 2 \omega_{\bf q} u_{\bf q} v_{\bf q} =0 
\end{eqnarray}

\noindent These identities are easily seen to hold if we use the 
following relations obeyed by the static quantities:

\begin{eqnarray}
&& u_{\bf q}^{2} = \frac{1}{2} \left( \frac{\tilde{e}_{\bf q}}
{\omega_{q}}+1 \right), \;\;\;\; v_{\bf q}^{2} = \frac{1}{2} 
\left( \frac{\tilde{e}_{\bf q}}{\omega_{\bf q}}-1 \right), \\
&& 2 u_{\bf q} v_{\bf q} = \frac{U_{pair}({\bf q})}{\omega_{\bf q}}.
\end{eqnarray}

In the equations of motion method the connection between  the
 Goldstone
 mode excitation operator
$\hat{{\cal Q}}_{G}$ and the number operator $\hat{N}$ goes as follows.
 Since  $[\hat{h},\hat{N}]=0$ one has

\begin{equation}
\langle \Phi | [\delta \hat{Q},[\hat{h},\hat{N}]] | \Phi \rangle =0.
\label{A}
\end{equation}

\noindent At first glance there is a difficulty to conclude from the 
above equation that $\hat{Q}_{G}$ is a zero-energy solution of 
Eq.(\ref{RPA}), caused by the presence of the term 
$\sum_{\bf k} ( u_{\bf k}^{2} + v_{\bf k}^{2}) \eta_{\bf k}^{\dagger}
 \eta_{\bf k}$ in Eq.(\ref{Nhat}) which does not belong to the 
general "ansatz" 
eq.(\ref{newquasiparticles}). However this term does not give 
any contribution to (\ref{A}) and since in our case the 
double-commutator is identical to the symmetrized 
double-commutator one has

\begin{equation}
\langle \Phi |[ \delta \hat{Q},\hat{h},\hat{Q}_{G}] | \Phi \rangle =0
\label{B}
\end{equation}
 
\noindent showing that the last term in Eq.(48) does not play any role
and indeed $\hat{Q}_{G}$ is a zero-energy solution of eq.(21). \\

In the HFB case we could proceed in the same fashion. However in 
this case the terms which do not belong to the HFB "ansatz" do 
contribute to the matrix element (25) and, as a consequence, 
the HFB equations do not have a zero-energy mode, the excitation 
spectrum always has a "gap".

\section{Numerical Results}

In this section, to ilustrate the predictions of the theory outlined
in this paper  and compare with the HFB and Bogoliubov approximations,
we present and discuss  the results obtained by solving  
equations (\ref{OmegaQ}) and {\ref{OmegaP} in a specific case. \\ 

To begin with, we choose as our pseudo-potential in momentum space, 
$\tilde{V}(q)$, 
a purely repulsive Gaussian defined by \cite{19,20} 

\begin{equation}
\tilde{V}(k) = \frac{4 \pi \hbar^{2} a}{m} e^{- \frac{\sigma^{2} k^{2}}{2}}
\end{equation}

\noindent where, as usual, the pseudo potential  at 
$k=0$ and the scattering length are related 
by the expression 

\begin{equation}
a = \frac{m \tilde{V}(0)}{4 \pi \hbar^{2}}
\end{equation}

We measure  energies in units of $\hbar^{2}/(2 m a^{2})$ and
 lengths in units of the scattering length $a$. The width of the 
pseudo potential was chosen to be of the order of the scattering 
length  $\sigma = 2.8 a$ as in \cite{20}. The only parameter 
left is the total density $\rho$ which was taken  
to be such that  $a^{3} \rho =  10^{-2}$ and $ 10^{-3}$. 
These choices of the ``dilution parameter'' 
falls between the values corresponding to the dilute experiments and 
to liquid Helium. Values such as $10^{-3}$ and $10^{-2}$ for 
the trapped BEC can be  achieved in experiments conducted 
close to a Feshbach resonance \cite{7}.

Once the parameters are specified,  we calculate the energy 
of the discrete branch,  the continuum threshold and  the structure 
of the excitation operators of the discrete branch. The first step in 
these calculations is to solve the self-consistent static
equations, (\ref{equilibrium1}),(\ref{equilibrium2}) and 
(\ref{constraint})which are needed in the determination of the 
${\cal A}_{+}$ and ${\cal A}_{-}$ matrices (\ref{A+}) and 
(\ref{A-}).The next step is to solve the coupled equations
(\ref{OmegaQ}) and (\ref{OmegaP}). The standard way to proceed is 
to take the thermodynamic limit  and solve the corresponding coupled 
integral equations. In this paper, we took a different route:  
we solved the coupled  equations (\ref{OmegaQ})and (\ref{OmegaP}),for 
a box with volume $V$.  The value of the volume $V$ is increased and 
the whole calculation is done again, until  
``saturation'' is observed, indicating  that the thermodynamic limit
 has been sufficiently  reached for these quantities.

In Fig.1 we present  results for
 $a^{3} \rho=10^{-2}$ since the qualitative behavior of
 the calculated quantities does 
not depend on the value of the density. We start by looking at 
the discrete branch in
Fig.1a. In the long-wave length limit this 
branch is gapless, as shown in section 3, has a phonon 
like dispersion relation i.e $\Omega_{P} = c P$ and it is always stable. 
 
We have also verified numerically that the continuum threshold starts 
at the minimum value for the two quasi-particles HFB energies, 
$\omega_{2}({\bf q},{\bf P})=\omega({\bf q}+{\bf P}/2) + 
\omega(-{\bf q}+{\bf P}/2)$ at a fixed value of ${\bf P}$, which 
in our case always happens at ${\bf q}=0$.

We can compare our results with the HFB and Bogoliubov
approximations. In the HFB approximation we have free quasi-particles,
and the one and two quasi-particle branches are decoupled. As
shown if Fig.1(b) both  branches  have a gap and, in the limit of long
wave length, it is not linear in ${\bf P}$. It is possible to show 
that these two
branches always cross at some value of ${\bf P}$. At small $P$  
the one quasi-particle branch energy is always lower than the two 
quasi-particle threshold due to the existence of the gap,
$\omega_{2}(0,0)=2 \omega(0)$, whereas for large $P$ is just the
 opposite, 
$\omega_{2}(0,{\bf P}) \approx P^{2}/2 < \omega({\bf P}) \approx P^{2}$,
 always leading to a crossing of the one quasi-particle branch and the
two quasi-particle threshold. As a consequence of the crossing the 
one quasi-particle branch always becomes unstable eventually. \\ 

For $a^{3} \rho=10^{-2}$, the crossing point happens at  $P=.63
a^{-1}$ 
therefore the one quasi-particle branch is stable for $P<.63 a^{-1}$, becoming 
unstable for $P>.63 a^{-1}$. If we compare with Fig. 1a
we see that the discrete branch "avoids" the crossing moving away from 
that point and rapidly approaching the two quasi-particle threshold
after the crossing point. This effect is seen in grater detail in
Fig. 2(c).

In the Bogoliubov approximation, shown in Fig. 1(c), the two branches 
are gapless and phonon like in the long wavelength limit. In this
approximation the one quasi-particle branch is always unstable
\cite{21}.For $a^{3} \rho=10^{-2}$ and  momenta $P < 0.54 a^{-1}$ the one 
quasi-particle branch and 
the two quasi-particle threshold  are degenerate. In this case the one
quasi-particle decays into two quasi-particles  where  one of them 
carries all the momentum and energy. This is possible because the one 
quasi-particle branch is gapless. For $P > .54 a^{-1}$ 
the one quasi-particle branch lies above the continuum threshold that 
occurs for zero relative momentum ${\bf q} =0$.

In conclusion, we found that the energy of the discrete branch, 
interpolates between the Bogoliubov one quasi-particle spectrum
 and the HFB two quasi-particle threshold, with a relatively sharp
transition region near the onset of instability of the HFB one
quasi-particle spectrum, as illustrated in Figs. 2(a), 2(c) for
$a^{3} \rho=10^{-2}$ and in Figs. 3(a), 3(c) for $a^{3} \rho=10^{-3}$.
 From these  graphs we also see
that the  sound velocities are  practically equal to the sound
velocities of the Bogoliuvov approximation.

We extracted  information on the composition of the excitation operator
  as a function of the total momentum ${\bf P}$ by 
calculating the quantity

\begin{equation}
c_{1}({\bf P}) = x_{\bf P}^{2} - y_{\bf P}^{2}
\end{equation}

\begin{figure}
\centering
\begin{minipage}[c]{.45\textwidth}
\centering
\caption{ The excitation spectrum according to the RPA, (a), to the
HFB theory, (b), and in the Bogoliubov approximation, (c) for 
$a^{3} \rho = 10^{-2}$. In fig 1(a) the solid lines indicate the
discrete branch and the continuum threshold, whereas in figs (b) and
(c) the lower solid line is the one quasi-particle branch. The dashed
line corresponds to the region where the energy of the  one
quasi-particle branch is greater than the threshold of the two
quasi-particle branch. The excitation energy is measured in units
of $\hbar^{2}/(2 m a^{2})$ and the momenta in units of $\hbar/a$. 
See text for details.}
\label{fig1}
\end{minipage}%
\hfill
\begin{minipage}[c]{.5\textwidth}
\begin{center}
\includegraphics[width=3.0in]{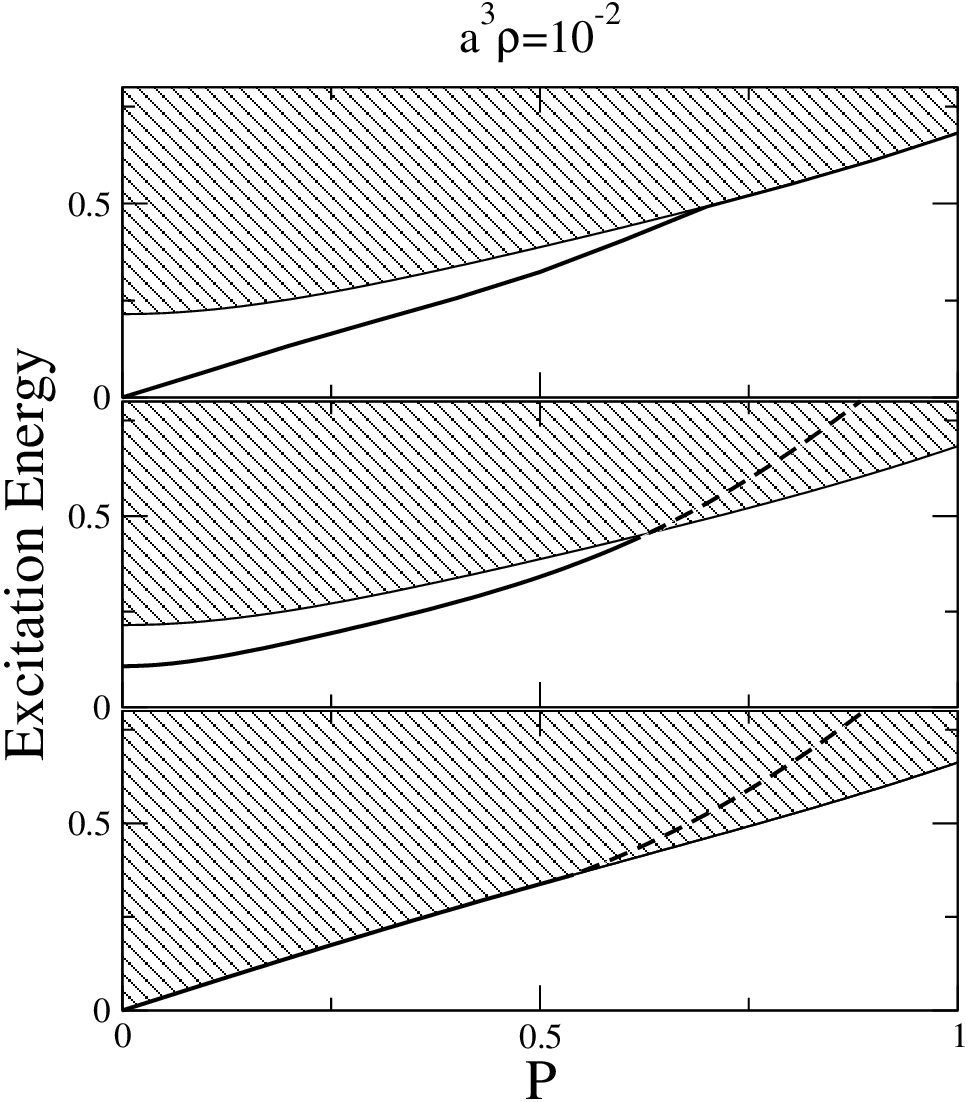}
\end{center}
\end{minipage}
\end{figure}

\noindent which corresponds  to the relative weight of the one 
quasi-particle component. Analogously for the two quasi-particle 
component we can define 

\begin{equation}
c_{2}({\bf P}) = \sum_{{\bf q} \ge 0} 
( X_{{\bf q},{\bf P}}^{2} - Y_{{\bf q},{\bf P}}^{2}). 
\end{equation}

\noindent These two relative weights  are related through the 
normalization condition (\ref{normalization}) that gives
\begin{equation}
c_{1}({\bf P}) + c_{2}({\bf P}) = 1
\end{equation}

Fig. 2(b) and 3(b) show the values of $c_{1}(P)$ and $c_{2}(P)$ as a 
function of the
total momentum $ P$ for $a^{3} \rho=10^{-2}$ and $10^-3$ respectively.
 In the long wavelength regime the excitation operator  
is predominantly a one quasi-particle operator.  On the other hand in 
the short wavelength regime it  becomes predominantly a two quasi-particle
operator as it approaches asymptotically the continuum threshold. Note
that there is a sharp transition between this two regimes, the effect
being more pronounced at higher densities. Comparing Figs. 2(a),2(b)
and 3(a),3(b), we see that the change from a
predominantly one quasi-particle to a predominantly two quasi-particle
physical content of the discrete branch of the excitation spectrum and
of its corresponding excitation operator occur in the same momentum range.

\begin{figure}
\centering
\begin{minipage}[c]{.45\textwidth}
\centering
\caption{Fig.2(a) shows the RPA discrete excitation energy and continuum
threshold (solid lines) and the Bogoliubov one quasi-particle
excitation energy (dot-dashed line). In Fig. 2(b) $c_{1}({\bf P})$ (solid line) and $c_{2}({\bf P})$ (dashed
line)  give , respectively, a measure
of the one and two quasi-particle character of the discrete branch, as
a function of $P$, for $a^{3}\rho= 10^{-2}$, Fig. 2(c) is a zoom of
Fig. 2(a) near the transition region. Units as in Fig.1. }
\label{fig2}
\end{minipage}%
\hfill
\begin{minipage}[c]{.5\textwidth}
\begin{center}
\includegraphics[width=3.0in]{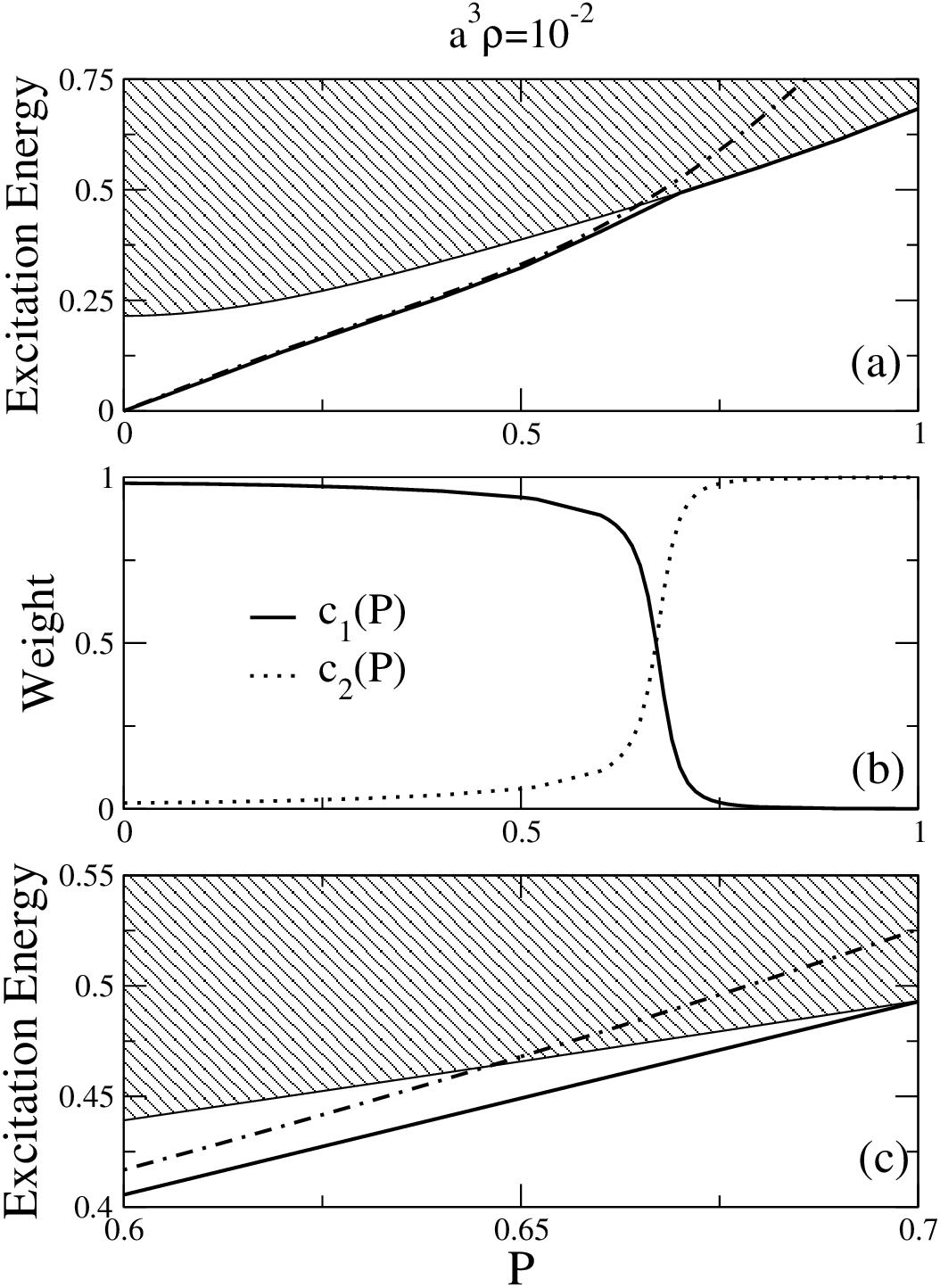}
\end{center}
\end{minipage}
\end{figure}

\begin{figure}
\centering
\begin{minipage}[c]{.45\textwidth}
\centering
\caption{Fig.3(a) shows the RPA discrete excitation energy and continuum
threshold (solid lines) and the Bogoliubov one quasi-particle
excitation energy (dot-dashed line). In Fig. 3(b) $c_{1}({\bf P})$ (solid line) and $c_{2}({\bf P})$ (dashed
line)  give , respectively, a measure
of the one and two quasi-particle character of the discrete branch, as
a function of $P$, for $a^{3}\rho= 10^{-3}$, Fig. 3(c) is a zoom of
Fig. 2(a) near the transition region. Units as in Fig.1. }
\label{fig3}
\end{minipage}%
\hfill
\begin{minipage}[c]{.5\textwidth}
\begin{center}
\includegraphics[width=3.0in]{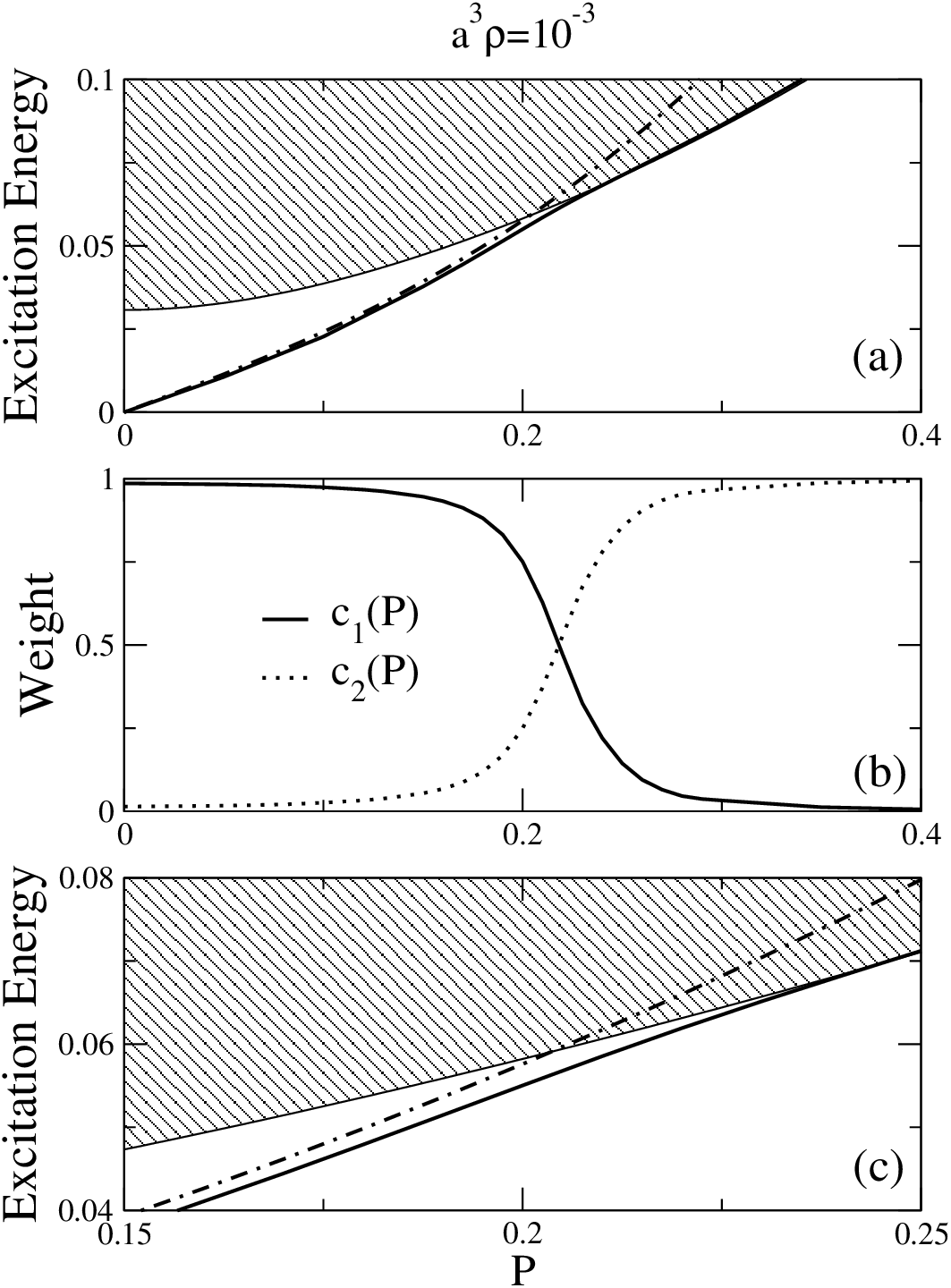}
\end{center}
\end{minipage}
\end{figure}

\section{Conclusion}

In this paper we presented a generalization of the HFB theory, cast in
terms of well known methods of equations of motion in order to access
the excitation spectrum of a condensed many- boson system. The key
ingredient for the generalization is the coupling of the one and two
quasi-particle components to form the excitation modes. These are
determined by solving the appropriate generalized Bogoliubov equations
for the relevant amplitudes and excitation energies. \\

The generalized Bogoliubov equations are shown to have a Goldstone
mode at zero transfered momentum, whose structure is related to that
generated by the particle number operator. Correspondingly, an
examination of the properties of the excitation spectrum reveals a
detached gapless excitation branch with phonon-like dispersion at small
momenta and a continuum branch starting at the two quasi-particle
threshold of the HFB theory. The detached branch has a predominantly
one quasi-particle character at small momenta, where it closely
approximates the Bogoliubov one quasi-particle spectrum. At high
momenta it approaches the continuum threshold and eventually acquires a
predominantly two quasi-particle character in what can be seen as an
avoided crossing situation, due to the coupling between one and two
quasi-particle components included in the calculation. The transition
from one to two quasi-particle character is relatively sharp and
occurs near the onset of instability of the HFB one quasi-particle
 spectrum. Differently from the Bogoliubov and HFB approximations the
detached phonon-like branch is always stable.  \\

The presence of the continuum branch serves, in particular, to give
physical significance to the HFB one quasi-particle energy gap, this
being the quantity which sets the appropriate energy scale for the
continuum threshold. \\

The features revealed in this generalized theory are closed linked to
the depletion of the condensate caused by the two body interaction
effects. They should be particularly relevant, therefore, in cases
where such depletion effect become important. This happens for larger
condensates densities and/or larger values of the relevant scattering
length (implying stronger effective interaction) a situation that may
be realized experimentally, for example, by taking advantages of
Feshbach resonances.  \\

\newpage
\appendix
\LARGE{{\bf Appendix A}} \normalsize 
\section{The Grand-Hamiltonian written in normal order}

The Grand-Hamiltonian can be written in the following form 

\begin{equation}
\hat{h} = \hat{h}_{0} + \hat{h}_{1} + \hat{h}_{2} + \hat{h}_{3} +\hat{h}_{4}
\label{A1}
\end{equation} 
\noindent where $\hat{h}_{i}$ corresponds to the normal 
ordered component  with $i$ quasi-particle 
operators.In turn,we will write each $\hat{h}_{i}$ as :

\begin{equation}
\hat{h}_{i}=\sum_{j=0}^{i} \hat{h}_{i-j,j}
\label{A1a}
\end{equation}
\noindent where $ i-j$ stands for the number of quasi-particle creation
operators and $ j $ for the number of annihilation operators.
In what follows we will give explicit expressions for all the 
components appearing in ({\ref{A1})  

\subsection{i) $\hat{h}_{0}$}
 
\begin{eqnarray}
\hat{h}_{0} &=& - z_{0}^{2} \mu + \frac{z_{0}^{4}}{2} V(0) + \sum_{\bf k} 
[e_{k} - \mu + ( V(0) + V(k)) z_{0}^{2}] v_{\bf k}^{2} \nonumber \\
&& - \sum_{\bf k} V(k) z_{0}^{2} u_{\bf k} v_{\bf k} + \frac{1}{2} 
\sum_{{\bf k_{1}},{\bf k_{2}}} (V(0) + V(|{\bf k_{1}} - {\bf k_{2}}|) 
v_{\bf k_{1}}^{2} v_{\bf k_{2}}^{2} \nonumber \\
&& + \frac{1}{2} \sum_{{\bf k_{1}},{\bf k_{2}}} V(|{\bf k_{1}}
 - {\bf k_{2}}|) u_{\bf k_{1}} v_{\bf k_{1}} u_{\bf k_{2}} v_{\bf k_{2}}
\label{A2}
\end{eqnarray} 

\subsection{ii) $\hat{h}_{1}$}

\begin{equation}
\hat{h}_{1}=h_{10}\eta_{0}^{\dagger}+h_{01}\eta_{0}
\label{A3}
\end{equation}
\noindent where
\begin{equation}
h_{10}=h_{01} = ( - \mu +  U_{h} +  U_{ex}^{nc}(0) +
U_{pair}^{nc}(0))z_{0}(x_{0}-y_{0})   
\label{A4}
\end{equation}

\noindent From eq.(\ref{A4}) it follows that $\hat{h}_{1}=0$ 
as a consequence of the 
equilibrium equation (\ref{equilibrium1}).

\subsection{iii) $\hat{h}_{2}$}
\begin{eqnarray}
\hat{h}_{2} &=& \sum_{k}h_{11}(\bf k) \eta_{{\bf k}}^{\dagger} \eta_{\bf k} 
\nonumber \\
&&+ h_{20}({\bf k})\eta_{\bf k}^{\dagger} \eta_{-\bf k}^{\dagger}
+h_{02}({\bf k}) \eta_{\bf k} \eta_{-\bf k}   
\label{A5}
\end{eqnarray}
\noindent where
\begin{equation}
h_{11}({\bf k})=\tilde{e}({\bf k}) \cosh 2 \sigma_{\bf k}  -  U_{pair}({\bf k})
\sinh 2 \sigma_{\bf k} 
\label{A5a}
\end{equation}
\begin{equation}
h_{20}({\bf k})=h_{02}({\bf k})=-\tilde{e}({\bf k}) \sinh 2 \sigma_{\bf k} +  
U_{pair}({\bf k}) \cosh 2 \sigma_{\bf k}
\label{A5b}
\end{equation}
\noindent From the equilibrium equation (\ref{equilibrium2}) we see
 immediately that the non-diagonal components in equation (\ref{A5}),
 $h_{20}({\bf k}),h_{02}({\bf k})$ vanish and the coefficient of the
diagonal component $ h_{11}({\bf k})$ is equal to $\omega_{{\bf k}}$ leading to
 
\begin{equation}
\hat{h}_{2}= \sum_{k} \omega({\bf k}) \eta_{\bf k}^{\dagger} \eta_{\bf
k} 
\label{A6}
\end{equation}

\subsection{iv) $\hat{h}_{3}$}
 
$\hat{h}_{3}$ can be written as

\begin{eqnarray}
\hat{h}_{3}& =& \sum_{{\bf k},{\bf k'}} \{ h_{1,2}({\bf k},{\bf k'}) \eta_{{\bf k}+{\bf k´}}^{\dagger} \eta_{\bf k} \eta_{\bf k'} + h_{2,1}({\bf k},{\bf k'}) \eta_{\bf k'}^{\dagger} \eta_{\bf k}^{\dagger} \eta_{{\bf k}+{\bf k'}} \nonumber \\
&& + h_{3,0} ({\bf k},{\bf k'})\eta_{{\bf k}+{\bf k'}}^{\dagger} \eta_{-{\bf k}}^{\dagger} \eta_{-{\bf k'}}^{\dagger} + h_{0,3}({\bf k},{\bf k'}) \eta_{{\bf k}+{\bf k'}} \eta_{-{\bf k}} \eta_{-{\bf k'}} \}  
\end{eqnarray}

\noindent where the $h_{i,k}$ coefficients are given by

\begin{eqnarray}
h_{1,2}({\bf k},{\bf k'}) &=&  h_{2,1}({\bf k},{\bf k'}) = \frac{z_{0}}{2} \left\{[u_{{\bf k}+{\bf k'}} u_{\bf k'}  + v_{{\bf k}+ \bf{k'}} v_{\bf k'}][u_{\bf k} - v_{\bf k}] V(k)  \right. \nonumber \\
&& \left. +[u_{{\bf k}+{\bf k'}} u_{\bf k}  + v_{{\bf k}+ \bf{k'}} v_{\bf k}][u_{\bf k'} - v_{\bf k'}] V(k')  \right. \nonumber \\ 
&&\left. -[u_{{\bf k}+{\bf k}'}- v_{{\bf k}+{\bf k}'}][ u_{\bf k} v_{\bf k'} + v_{\bf k} u_{\bf k'}] V(|{\bf k}+{\bf k'}|) \right\} \label{lala} \\
h_{3,0}({\bf k},{\bf k'})  &=& h_{0,3}({\bf k},{\bf k'}) =\frac{z_{0}}{2} \left\{[u_{{\bf k}+{\bf k'}} v_{\bf k} v_{\bf k'} - v_{{\bf k}+{\bf k'}} u_{\bf k} u_{\bf k'}][V(k) + V(k')] \right\}. \nonumber
\end{eqnarray}

\noindent Note that all the coefficients obey the symmetry propriety $h_{i,j}({\bf k},{\bf k'}) = 
h_{i,j}(-{\bf k},-{\bf k'})=h_{i,j}({\bf k'},{\bf k})$.

\subsection{v) $\hat{h}_{4}$ } 

\noindent Analogously to the previous case we can write $\hat{h}_{4}$ as

\begin{eqnarray}
\hat{h}_{4} &=& \sum_{{\bf k_{1}},{\bf k_{2}},{\bf q}} h_{1,3}({\bf k_{1}},{\bf k_{2}},{\bf q})  \eta_{{\bf k_{2}}-{\bf q}}^{\dagger} \eta_{-{\bf k_{1}}-{\bf q}} \eta_{{\bf k_{1}}} \eta_{\bf k_{2}} \\  
&& +  h_{3,1}({\bf k_{1}},{\bf k_{2}},{\bf q}) \eta_{{\bf
k_{1}}}^{\dagger} \eta_{{\bf k_{2}}}^{\dagger} \eta_{-{\bf k_{1}}-{\bf
q}}^{\dagger}  \eta_{{\bf k_{2}}-{\bf q}} \nonumber \\
&&+ h_{4,0}({\bf k_{1}},{\bf k_{2}},{\bf q}) \eta_{{\bf k_{1}}+{\bf q}}^{\dagger} \eta_{{\bf k_{2}}-{\bf q}}^{\dagger} \eta_{-{\bf k_{1}}}^{\dagger}  \eta_{-{\bf k_{2}}}^{\dagger} \nonumber \\
&& + h_{0,4}({\bf k_{1}},{\bf k_{2}},{\bf q})\eta_{{\bf k_{1}}+{\bf q}}  \eta_{{\bf k_{2}}-{\bf q}} \eta_{-{\bf k_{1}}}  \eta_{-{\bf k_{2}}} \nonumber \\
&&+ h_{2,2} ({\bf k_{1}},{\bf k_{2}},{\bf q})\eta_{{\bf k_{1}}+{\bf q}}^{\dagger} \eta_{{\bf k_{2}}-{\bf q}}^{\dagger}  \eta_{{\bf k_{1}}}  \eta_{\bf k_{2}}  
\end{eqnarray}

\noindent where the coefficients are given by

\begin{eqnarray}
&& h_{3,1}({\bf k_{1}},{\bf k_{2}},{\bf q}) = h_{1,3} ({\bf k_{1}},{\bf k_{2}},{\bf q}) =- [ u_{{\bf k_{2}}-{\bf q}} v_{{\bf k_{1}}+{\bf q}} u_{\bf k_{1}} u_{\bf k_{2}} + \nonumber \\
&& v_{{\bf k_{2}}-{\bf q}} u_{{\bf k_{1}}+{\bf q}} v_{\bf k_{1}} v_{\bf k_{2}}] V(q)  \\
&&\hat{h}_{4,0}({\bf k_{1}},{\bf k_{2}},{\bf q}) =  \hat{h}_{0,4} ({\bf k_{1}},{\bf k_{2}},{\bf q}) =\frac{1}{2} u_{{\bf k_{1}}+{\bf q}} u_{{\bf k_{2}}-{\bf q}} v_{\bf k_{1}} v_{\bf k_{2}} V(q) \\
&& \hat{h}_{2,2}({\bf k_{1}},{\bf k_{2}},{\bf q}) =  \frac{1}{2} \{[u_{{\bf k_{1}} + {\bf q}} u_{\bf k_{1}} + v_{{\bf k_{1}}+ {\bf q}} v_{\bf k_{1}}][u_{{\bf k_{2}}-{\bf q}} u_{\bf k_{2}} +  v_{{\bf k_{2}}-{\bf q}} v_{\bf k_{2}}]V(q) \nonumber \\
&&+ [ u_{{\bf k_{2}}-{\bf q}} u_{\bf k_{1}} v_{{\bf k_{1}}+{\bf q}} v_{\bf k_{2}}+ 
u_{{\bf k_{1}}+{\bf q}} u_{{\bf k_{2}}} v_{{\bf k_{2}}-{\bf q}} v_{\bf
k_{1}}] V(|{\bf k_{1}} + {\bf k_{2}}|) \}  
\end{eqnarray}
These coefficients,except $ h_{3,1}$ and $h_{1,3}$ obey the symmetry property  
\begin{equation}
h_{i,j}({\bf k_{1}}, {\bf k_{2}}, {\bf q}) = h_{i,j}({\bf k_{2}}, {\bf
k_{1}}, -{\bf q})= h_{i,j}(-{\bf k_{1}}, -{\bf k_{2}},- {\bf q})
\end{equation}

\newpage

\appendix
\LARGE{{\bf Appendix B}} \normalsize 
\section{RPA Matrices}

In this appendix we evaluate the elements of the matrices $\cal A$
and $\cal B$. The key property 
that we will use to calculate the average value of the symmetrized 
double commutators is that $| \Phi \rangle$ is the quasi-particle vacuum. In
order to work with more compact expressions we introduce the following
quantities 
\begin{eqnarray}
C_{0}({\bf k}) &=& u_{\bf k} - v_{\bf k} \\
C_{1}({\bf k},{\bf k'}) &=& C_{1}({\bf k}',{\bf k}) = C_{1}(-{\bf k},-{\bf k'}) =u_{\bf k} u_{\bf k'} + v_{\bf k} v_{\bf k'} \label{c1} \\
C_{2}({\bf k},{\bf k'}) &=& C_{2}({\bf k}',{\bf k}) = C_{2}(-{\bf k},-{\bf k'}) =u_{\bf k} v_{\bf k'} + v_{\bf k} u_{\bf k'}. \label{c2}
\end{eqnarray}
\noindent where (\ref{c1}) and (\ref{c2}) are usually called coherence
 factors  and are well known in the study of Bose systems \cite{9}.

\subsection{The matrix $\cal A$}

${\cal A}^{11}({\bf P})$ is given by, 
\begin{equation}
{\cal A}^{11}({\bf P}) = \langle \Phi | [ \eta_{\bf P}, H, 
\eta_{\bf P}^{\dagger}] | \Phi \rangle. 
\end{equation}
\noindent From the property that $| \Phi \rangle$ is the
quasi-particle vacuum it follows immediately
that only $\hat{h}_{2}$ contributes to this matrix element, with the result
  
\begin{equation}
{\cal A}^{11}({\bf P}) = \omega({\bf P}). 
\end{equation}

\noindent The elements of the $1 \times npair$ matrix  ${\cal A}^{12}$
 are given by

\begin{equation}
{\cal A}^{12}({\bf q};{\bf P}) =  \langle \Phi | [ \eta_{\bf P}, H,  
\frac{\eta_{{\bf q} + 
{\bf P}/2}^{\dagger} \eta_{-{\bf q} + {\bf P}/2}^{\dagger}}{\sqrt{1+ 
\delta_{{\bf q},0}}}  ] | \Phi \rangle
\end{equation}

\noindent  The only contribution to this matrix element comes from 
$\hat{h}_{3}$ through the term  $\hat{h}_{12}$ , giving

\begin{eqnarray}
{\cal A}^{12}({\bf q};{\bf P}) &=&- 2 \frac{h_{1,2}({\bf q}+{\bf P}/2,-{\bf q}+{\bf P}/2)}{\sqrt{1+ \delta_{{\bf q},0}}}.
\label{A12}
\end{eqnarray}
\noindent Using the expression for $h_{12}$ term  show in Eq.(\ref{lala}) 
we can write (\ref{A12}) in terms of the coherence factors (\ref{c1}),
(\ref{c2}}) obtaining

\begin{eqnarray}
&& {\cal A} ^{12}({\bf q};{\bf P}) = \frac{1}{\sqrt{1+ \delta_{{\bf
q},0}}} 
\{ C_{0}(+) C_{1}({\bf P},-) V(+)
+ C_{0}(-) C_{1}({\bf P},+) V(-)  \nonumber \\
&&  - C_{0}(P) C_{2}(+,-) V(P) \}
\end{eqnarray}

\noindent where we used the notation $\pm =\pm {\bf q} + {\bf P}/2$
and $\pm' = \pm {\bf q}' + {\bf P}/2$. 
\noindent The elements of the  $npairs \times 1$ matrix   ${\cal A}^{21}$
is the hermitian conjugate of the matrix ${\cal A}^{12}$.
\noindent The elements of $npairs \times npairs$ matrix 
${\cal A}^{22}({\bf q'},{\bf q};{\bf P})$ are given by  

\begin{equation}
{\cal A}^{22}({\bf q'},{\bf q};{\bf P}) = \langle \Phi
 |[ \frac{\eta_{{\bf q'} + 
{\bf P}/2} \eta_{-{\bf q'} + {\bf P}/2}}{\sqrt{1+ \delta_{{\bf q'},0}}}, H , \frac{\eta_{{\bf q} + 
{\bf P}/2}^{\dagger} \eta_{-{\bf q} + {\bf P}/2}^{\dagger}}{\sqrt{1+ \delta_{{\bf q},0}}}] | \Phi \rangle,
\end{equation}
\noindent Both $\hat{h}_{2}$, through the term $\hat{h}_{1,1}$, and 
$\hat{h}_{4}$ through the term $\hat{h}_{2,2}$ contribute
to this matrix element, leading to 
 
\begin{eqnarray}
&& {\cal A}^{22}({\bf q'},{\bf q};{\bf P}) =  [\omega(-{\bf q}+{\bf P}/2) + \omega({\bf q}+{\bf P}/2)] \delta_{{\bf q},{\bf q}'} +
\frac{1}{\sqrt{(1 + \delta_{{\bf q},0})(1 + \delta_{{\bf q}',0})}} \{ \nonumber \\
&& h_{2,2}({\bf q}+{\bf P}/2,-{\bf q}+{\bf P}/2,{\bf q}'-{\bf q})  
+ h_{2,2}({\bf q}'+{\bf P}/2,-{\bf q}'+{\bf P}/2,{\bf q}-{\bf q}') 
+ \nonumber \\
&& h_{2,2}(-{\bf q}+{\bf P}/2,{\bf q}+{\bf P}/2,{\bf q}'+{\bf q})  
+ h_{2,2}(-{\bf q}'+{\bf P}/2,{\bf q}'+{\bf P}/2,{\bf q}+{\bf q}') \}.  
\end{eqnarray}

\noindent Using the expression for $h_{2,2}$ calculated in the
previous appendix we get

\begin{eqnarray}
&& {\cal A}^{22}({\bf q'},{\bf q};{\bf P}) =  [\omega(-) + \omega(+)] \delta_{{\bf q},{\bf q}'} + 
\frac{1}{\sqrt{(1 + \delta_{{\bf q},0})(1 + \delta_{{\bf q}',0})}} \{  \nonumber \\ 
&& C_{2}(+',-') C_{2}(+,-) V(P) + C_{1}(+,+') C_{1}(-,-') V(|{\bf q}-{\bf q}'|) + \nonumber \\
&& C_{1}(+',-) C_{1}(+,-') V(|{\bf q}+{\bf q}'|) \}
\end{eqnarray}

\subsection{The matrix $\cal B$ }

For the matrix {\cal B}  we start with  $B^{11}$

\begin{equation}
{\cal B}^{11}({\bf P}) = \langle \Phi | [ \eta_{\bf P}, H, \eta_{\bf -P}] | \Phi \rangle,
\end{equation}
\noindent By inspection, we see    that there are no terms in the 
Hamiltonian that contributes to this matrix element, hence
\begin{equation}
{\cal B}^{11}({\bf P}) =0.
\end{equation}
\noindent Next we consider the matrix elements of  ${\cal B}^{12} $
given by
\begin{equation} 
{\cal B}^{12}({\bf q}; {\bf P}) = \langle \Phi | [ \eta_{\bf P}, H,  \frac{\eta_{-{\bf q} - 
{\bf P}/2} \eta_{{\bf q} - {\bf P}/2}}{\sqrt{1+ \delta_{{\bf q},0}}}  ] | \Phi \rangle,
\end{equation}  
\noindent In this case the  only contribution comes from $\hat{h}_{3}$ through the term $\hat{h}_{3,0}$ 
\begin{eqnarray} 
{\cal B}^{12}({\bf q}; {\bf P}) &=& 
 - \frac{2}{\sqrt{1+ \delta_{{\bf q},0}}}[     
 h_{3,0}({\bf q}+{\bf P}/2,-{\bf q}+{\bf P}/2) + \nonumber \\
&& h_{3,0}(-{\bf q}+{\bf P}/2,-{\bf P}) + h_{3,0}({\bf q}+{\bf
 P}/2,-{\bf P}) ]  
\end{eqnarray}
\noindent which can be written in terms of the coherence factors as 
\begin{eqnarray}
&& {\cal B}^{12}({\bf q}; {\bf P}) = -\frac{1}{\sqrt{1+\delta_{{\bf q},0}}}\{C_{0}(+) 
C_{2}({\bf P},-) V(+) \nonumber \\
&& C_{0}(-) C_{2}({\bf P},+) V(-) +C_{0}(P)C_{2}(+,-) V(P)] \} 
\end{eqnarray}
The matrix $ {\cal B}^{12}$ is the transpose of $ {\cal B}^{12}$. \\

For $B^{22}({\bf q'},{\bf q};{\bf P})$ we have 
\begin{equation}
B^{22}({\bf q'},{\bf q};{\bf P}) = \langle \Phi |[ \frac{\eta_{{\bf q'} + 
{\bf P}/2} \eta_{-{\bf q'} + {\bf P}/2}}{\sqrt{1+ \delta_{{\bf q'},0}}}, H 
 , \frac{\eta_{{\bf q} - 
{\bf P}/2} \eta_{-{\bf q} - {\bf P}/2}}{\sqrt{1+ \delta_{{\bf q},0}}} | \Phi \rangle,
\end{equation}
\noindent  The only non vanishing contribution to this matrix element  comes 
from $\hat{h}_{4}$ through the term $\hat{h}_{4,0}$, giving:

\begin{eqnarray}
&& B^{22}({\bf q'},{\bf q};{\bf P})= -h_{0,4}(-{\bf q}+{\bf P}/2,-{\bf q}'-{\bf P}/2,{\bf q}-{\bf q'}) \nonumber \\
&& -h_{0,4}({\bf q}+{\bf P}/2,-{\bf q'}-{\bf P}/2,-{\bf q}'-{\bf q}) \nonumber \\
&& -h_{0,4}(-{\bf q}+{\bf P}/2,{\bf q}+{\bf P}/2,-{\bf q}'+{\bf q})\nonumber \\
&& -h_{0,4}({\bf q}+{\bf P}/2,-{\bf q}+{\bf P}/2,-{\bf q}-{\bf q}')\nonumber \\
&&  -h_{0,4}(-{\bf q}+{\bf P}/2,{\bf q'}-{\bf P}/2,{\bf q}'+{\bf q})  \nonumber \\
&& -h_{0,4}({\bf q}+{\bf P}/2,{\bf q}'-{\bf P}/2,{\bf q}'-{\bf q})\nonumber \\
&&  -h_{0,4}(-{\bf q}+{\bf P}/2,{\bf q}+{\bf P}/2,{\bf q}+{\bf q'}) \nonumber \\
&&- h_{0,4}({\bf q}+{\bf P}/2,-{\bf q}+{\bf P}/2,{\bf q}'-{\bf q}) \nonumber \\
&&  -h_{0,4}({\bf q}'-{\bf P}/2,-{\bf q}+{\bf P}/2,{\bf P}) -h_{0,4}({\bf q}'-{\bf P}/2,{\bf q}+{\bf P}/2,{\bf P})    \nonumber \\
&&  -h_{0,4}(-{\bf q}'-{\bf P}/2,-{\bf q}+{\bf P}/2,{\bf P}) -
h_{0,4}(-{\bf q}'-{\bf P}/2,{\bf q}+{\bf P}/2,{\bf P}) \nonumber \\ 
&& + q \rightleftharpoons q',P \rightarrow -P  
\end{eqnarray}

\noindent which can be written, in terms of the coherence factors as

\begin{eqnarray}
&& B^{22}({\bf q'},{\bf q};{\bf P})= - \frac{1}{\sqrt{(1 + \delta_{{\bf q},0})(1 + \delta_{{\bf q}',0})}} \{ C_{2}[+,+'] C_{2}[-,-'][V(|{\bf q} + {\bf q}'|)  \nonumber \\
&& + V(|{\bf q} - {\bf q}'|)] + C_{2}[+,-] C_{2}[+',-'] V(P) \} 
\end{eqnarray}

\section{Acknoledgements}

This work was supported by FAPESP under the contract 
number 00/00649-9. P.T. and M.O.C.P. are supported by FAPESP. EJVP is
partially supported by CNPq. The authors would like to thank Dr. Eddy
Timmermnas for his critical reading of the manuscript.

\end{document}